\documentclass[epsfig,aps,prl,twocolumn,amsmath,amssymb]{revtex4}

\usepackage{dcolumn}
\usepackage{bm}
\usepackage{graphicx}

\usepackage[english]{babel}

\begin{document}

\title{Inhibited spontaneous emission of quantum dots observed in a 3D photonic band gap}

\author{M.D. Leistikow$^{1,2}$}

\author{A.P. Mosk$^{1}$}

\author{S.R. Huisman$^{1}$}

\author{A. Lagendijk$^{1,2}$}

\author{W.L. Vos$^{1}$}
\altaffiliation{Email: w.l.vos@utwente.nl, www.photonicbandgaps.com}

\affiliation{$^1$Complex Photonic Systems (COPS), MESA+ Institute for Nanotechnology, University of Twente, The Netherlands}

\affiliation{$^2$FOM Institute for Atomic and Molecular Physics (AMOLF), 1098 XG Amsterdam, The Netherlands}


\date{Prepared May 23rd, 2011}

\begin{abstract}
We present time-resolved emission experiments of semiconductor quantum dots in silicon 3D inverse-woodpile photonic band gap crystals.
A systematic study is made of crystals with a range of pore radii to tune the band gap relative to the emission frequency.
The decay rates averaged over all dipole orientations are inhibited by up to a factor of $12 \times$ in the photonic band gap, and enhanced up to $2 \times$ outside the gap, in quantitative agreement with theory.
We discuss the effects of spatial inhomogeneity, non-zero non-radiative decay, and transition dipole orientations on the observed inhibition in the band gap.
\end{abstract}
\maketitle

In the field of cavity quantum electrodynamics (QED) it has been recognized that the nano-environment of a two-level quantum system may serve to tailor the fundamental light-matter interactions~\cite{Haroche92,Scully01}.
Of particular interest is the broadband and radical suppression of vacuum fluctuations in a three-dimensional (3D) photonic band gap, \emph{i.e.}, a frequency range for which light is forbidden for all wave vectors and all polarizations~\cite{Yablonovitch87}.
Such band gaps are expected in 3D photonic crystals, \emph{i.e.}, dielectric nanostructures with periodicities less than half the optical wavelength~\cite{Joannopoulos08}.
Anticipated cavity QED effects of band gaps include complete inhibition of spontaneous emission, photon-atom bound states, collective laser-like emission, and intriguing fractional decay~\cite{John90,Li01,Vats02,Wang03,Kristensen08}.

To date, the cavity QED effects of 3D photonic band gaps on two-level light sources have only been studied in theory~\cite{John90,Li01,Vats02,Wang03,Kristensen08}.
In these studies, one usually considers a single two-level source (or N identical ones) with ideal $100\%$ quantum efficiency that is excited once at t=0, and that is placed in a perfect and infinitely extended photonic crystal.
Clearly, this idealized situation differs from real situations.
Real sources are inhomogeneously broadened, have non-zero non-radiative decay, are repeatedly excited in time, and are embedded in photonic crystals of finite extent.
In this paper, we perform a first experimental study of cavity QED effects of real 3D photonic band gap crystals on embedded two-level light sources.
To this end, we have developed Si photonic crystals with a diamond-like structure that have broad gaps~\cite{Ho94}.
We perform time-correlated single photon counting experiments on embedded semiconductor quantum dots and observe broadband inhibition inside and enhancement outside the band gap.
We notably find that the effects of spatial inhomogeneity, non-zero non-radiative decay, and transition dipole orientations are relevant in the band gap.
Our results allow for the first time to test predictions from several different theoretical models.

\begin{figure}
\centering
\includegraphics[width=\columnwidth]{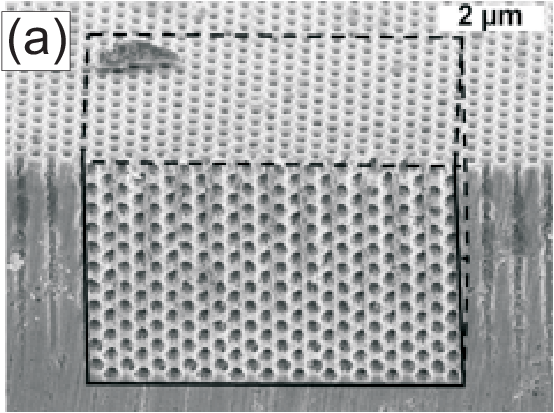}
\includegraphics[width=\columnwidth]{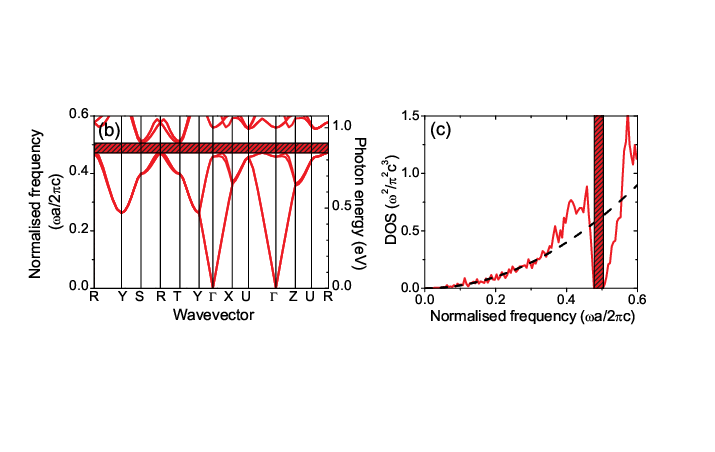}
\caption{(color)
(a) Scanning electron micrograph of a 3D inverse woodpile photonic crystal made from silicon.
The crystal consists of two perpendicularly etched sets of carefully aligned pores, and is delimited by the dashed lines.
(b) Band structures of an infinite inverse woodpile photonic crystal with pore radius $170$ nm, 
using $\epsilon=12.1$ for silicon and $\epsilon=2.25$ for toluene-filled pores.
The band gap is indicated with the red bar.
Inset: first Brillouin zone.
(c) Density of states (DOS) per volume for the same crystal calculated with 2000 $k$-points~\cite{Johnson01}.
The DOS vanishes in the band gap.
Dashed curve: quadratic behavior in the low frequency limit.
}
\label{SEMbandstructureDOS}
\end{figure}

Silicon 3D inverse woodpile photonic band gap crystals were fabricated by a CMOS-compatible method described in Ref.~\cite{Woldering08}.
In brief, two orthogonal sets of pores are etched consecutively in a silicon wafer by reactive ion etching after careful alignment.
Fig.~\ref{SEMbandstructureDOS}(a) shows a scanning electron micrograph of a typical crystal.
The cubic crystals have lattice parameters $a = 693$ and $c = 488$ nm ($a/c = \sqrt2$) and a range of pore radii ($136 < r < 186$ nm) to tune the band gap relative to the emission spectrum.
The 3D crystal extends over $L^{3} = 12 \times 12 \times 12~ac^{2}$, which exceeds the Bragg length in every direction~\cite{Huisman11}.
The good optical quality and high photonic strength of our crystals have been confirmed by optical reflectivity where intense and broad peaks are seen.
The stopbands overlap for all probed directions and polarizations, which is an experimental signature of a photonic band gap~\cite{Huisman11}.

Fig.~\ref{SEMbandstructureDOS}(b) shows the calculated band structure~\cite{Johnson01} for one of our inverse woodpile crystals.
A broad band gap appears in the frequency range where modes are forbidden for all wavevectors.
Fig.~\ref{SEMbandstructureDOS}(c) shows the corresponding density of states (DOS).
At low frequency the DOS increases quadratically as is typical of an effective medium.
Beyond $0.35$ the photonic regime is entered as the DOS deviates from the parabola.
The DOS completely vanishes in the photonic band gap between 0.478 and 0.504.

\begin{figure}
\centering
\includegraphics[width=0.9\columnwidth]{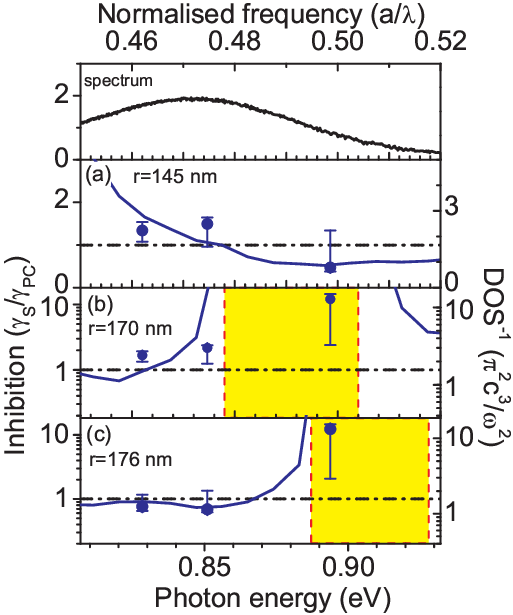}
\caption{(a) to (c): measured inhibition $\gamma_{S}/\gamma_{PC}$ (open circles) versus photon energy for crystals with indicated pore radii.
Each vertical bar spans variations of 2 to 3 measurements taken on a same crystal on different days.
The horizontal lines indicate reference level 1.
Circles represent extremal data since a lower inhibition is likely biased by dots outside the crystal.
The calculated inverse density of states (DOS) are shown as curves.
The top panel shows the quantum dot emission spectrum.}
\label{ratevsfrequency}
\end{figure}

To study emission from two-level sources, we immersed the crystals in suspensions of PbS colloidal quantum dots in toluene (Evident).
The concentration was as low as $2 \cdot 10^{-6}$ M to avoid dot-dot interactions such as energy transfer.
The dots emit at photon energies between 0.8 and 0.9 eV including the telecom range (Fig.~\ref{ratevsfrequency}), and their transition dipole orientations sample all directions.
We measured room temperature decay rates at three energies in the quantum dot spectrum: 0.828, 0.850, and 0.893 eV ($\lambda = 1500$, 1460, 1390 nm respectively).
Outside the crystals the dots reveal single exponential decay with an energy-independent rate. 
By keeping the dots in suspension the photophysical properties remain stable for months, much longer than the short stability of dried dots in previous studies~\cite{Lodahl04}.

\begin{figure}
\centering
\includegraphics[width=0.9\columnwidth]{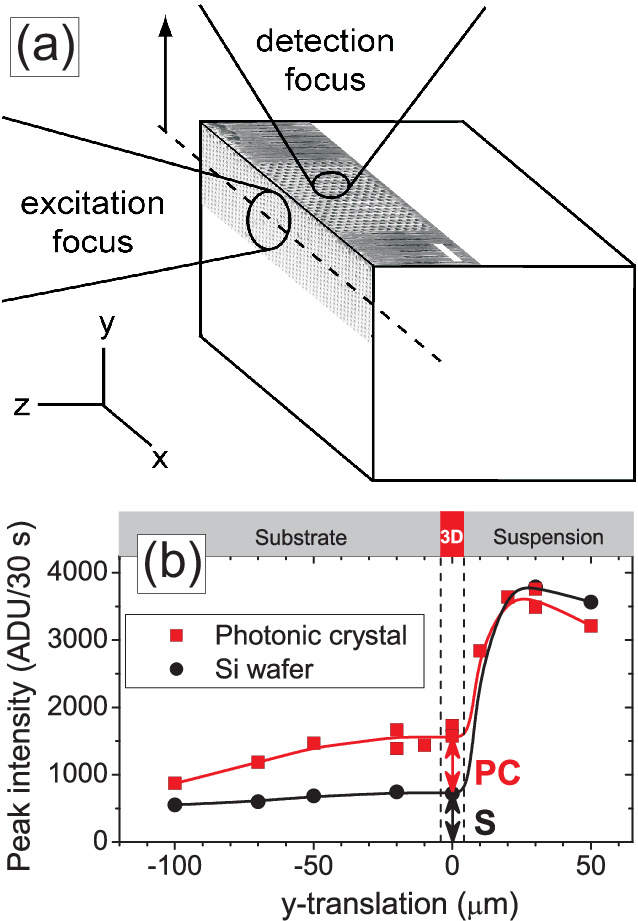}
\caption{
(a) Drawing of the crystal position relative to excitation and and detection foci.
The coordinate system of the position scans is shown.
(b) Peak intensity of quantum dots at 0.842 eV (1475 nm) vs. y-position in an inverse woodpile crystal (red squares) and near a silicon wafer (black circles), curves are guides to the eye.
The photonic crystal reveals twice as much signal as bulk silicon.
The measured intensities ($PC, S$) are input for the time-resolved emission model.}
\label{scanningexcfocus}
\end{figure}

The quantum dots were excited with short (11 ps) light pulses from a frequency-doubled Nd$^{3+}$:YAG laser (Time Bandwidth Cougar)
at $\lambda=532$ nm at a repetition rate of 409 kHz.
The laser power was kept sufficiently low to avoid saturation and ensure that the dots remain in the linear regime.
Light collected from the quantum dots was resolved by a monochromator (Acton 2500i)
set to 0.003 eV bandwidth, sufficiently narrow to resolve the photonic gaps.
The excitation light was focused with an NA=0.12; $4\times$ objective on the crystal, see Fig.~\ref{scanningexcfocus}(a).
To minimize background from the suspension, light from the quantum dots was collected at $90^{\circ}$ from the excitation beam by a detection objective with NA=0.7; $100\times$.
To precisely align the crystal the excitation focus was scanned along the edge of the structure.
Fig.~\ref{scanningexcfocus}(b) shows a scan of intensity versus y-position of the excitation focus.
We see a sharp transition from high (in suspension) to lower intensity in the crystal.
In a scan performed on bulk silicon next to the crystal a sharp edge is also seen.
Here the intensity is about half of the intensity collected from the photonic crystal.
Hence about half of the intensity is emitted by quantum dots in the photonic crystal while the other half stems from quantum dots in suspension.
Therefore, we conclude that a significant spontaneous emission signal originates from quantum dots in our photonic band gap crystals.

\begin{figure}
\centering
\includegraphics[width=0.9\columnwidth]{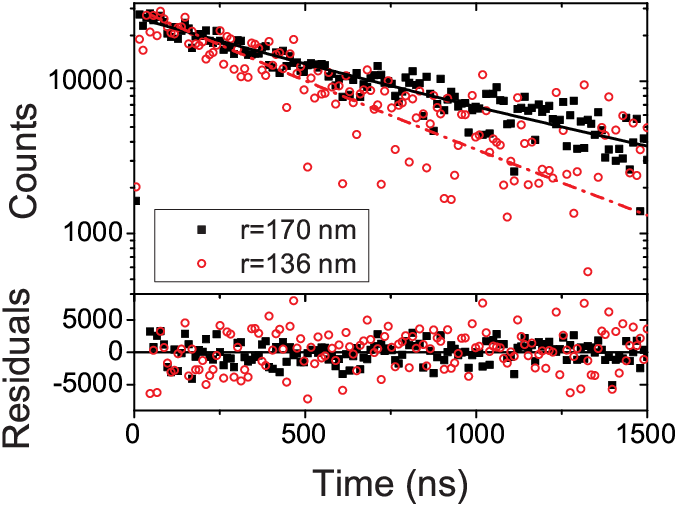}
\caption{Time-resolved spontaneous emission measured at 0.850 eV on two different photonic crystals with pore radii
170 nm (closed squares) and 136 nm (open circles).
Bi-exponential models (Eq.~\ref{eq:fitting}) are indicated by the curves (solid for 170 nm, dash-dotted for 136 nm).
Bottom panel: the residuals are random indicating high quality fits. }
\label{decaycurve}
\end{figure}

To precisely measure the arrival time of the emitted photons, time-correlated single photon counting was employed with a cooled photomultiplier tube (Hamamatsu NIR) and a timing card (Picoquant PicoHarp 300).
Since we perform experiments at relatively low photon energy in the near infrared to avoid Si absorption, the detector darkcount is substantial ($3 \cdot 10^5$ counts/s) and the quantum dot signal is relatively low ($< 1 \%$ of the repetition frequency, or $<4 \cdot 10^3$ counts/s).
Therefore, we have collected for long times (hours) to obtain sufficient statistics, and carefully subtracted the background~\cite{footnote:background}.
Fig.~\ref{decaycurve} shows time-resolved spontaneous emission for quantum dots in two different photonic crystals with pore radii $170$ and $136$ nm. The emission inside the bandgap ($r=170$ nm) decays slower than outside the bandgap ($r=136$ nm), which confirms that the excited state lifetime of quantum dots is controlled by the photonic crystals.

Interestingly, Fig.~\ref{decaycurve} reveals that the photon arrival times decay monotonously in time to a vanishing intensity, in agreement with Ref.~\cite{Li01}; we do not detect signatures of fractional decay or oscillations predicted in Refs.~\cite{Vats02,Wang03,Kristensen08}.
A possible reason for this discrepancy may be that those predicted features are not robust to ensemble averaging.
Other reasons could be the finite dynamic range (1.5 decades) in our experiments, the assumption of very large oscillator strengths in some models, or that theories predict excited-state population dynamics instead of photon arrival times while these two phenomena do not necessarily have the same dynamics, see Ref.~\cite{vanDriel07}.

Since the emission originates both from quantum dots in the crystal and in suspension, we model the time-resolved emission with a double-exponential:
\begin{equation}
f(t) = I (S \exp(- \gamma_{S} t) + PC \exp(-\gamma_{PC} t)).
\label{eq:fitting}
\end{equation}
Here $\gamma_{PC}$ is the rate of quantum dots in the photonic crystal that we wish to obtain, and $\gamma_{S}$ is the rate of  dots outside the crystal that is separately found to be $\gamma_{s} = 1.9~\mu s^{-1}$.
$S$ and $PC$ are determined by scans of the detection focus (see Fig.~\ref{scanningexcfocus}(b)).
The high quality of our model is confirmed by residuals randomly distributed about 0 (Fig.~\ref{decaycurve} bottom), and by the goodness of fit $\chi^2_{red}$ of 1.01 and 0.93 near unity.
From Fig.~\ref{decaycurve} we obtain decay rates $\gamma_{PC} = 3.1~\mu s^{-1}$ and $0.88~\mu s^{-1}$ for two different crystals.
Therefore these data reveal respectively enhanced and inhibited emission compared to quantum dots in suspension.

Fig.~\ref{ratevsfrequency} (a) to (c) collects the decay rates measured on several crystals with a range of pore radii and hence gap frequencies.
To investigate the inhibition in the photonic band gap range we plot the ratio $\gamma_{S}/\gamma_{PC}$.
For all crystals a good agreement between the measured inhibition and the inverse DOS is observed.
The agreement is much better than previously found in inverse opals~\cite{Lodahl04}, likely since our quantum dots occupy a much larger fraction of the 
unit cell thus providing a better average of the local density of states (LDOS)~\cite{Sprik96}.
For the $145$ nm crystal (Fig.~\ref{ratevsfrequency}(a)) the photon energy is higher than the band gap frequency.
An enhanced decay rate up to $2 \times$ is observed with increasing photon energy, in agreement with the DOS.
The $170$ nm crystal (Fig.~\ref{ratevsfrequency}(b)) reveals up to a 2-fold inhibition at frequencies just below the gap.
Strikingly, a strong up to $12 \times$ inhibition is found at photon energies deep in the band gap where a minimum decay rate of $0.16~\mu s^{-1}$ is observed.
The corresponding lifetimes up to $T_{1} = 6.25~\mu$s are extremely long for quantum dots, and confirm strongly shielded vacuum fluctuations, which is favorable for applications in quantum information processing~\cite{Imamoglu99}.
The occurrence of a lower inhibition in the band gap may be caused by slight misalignments that bias results to $\gamma_{S}$; since the measurements take long it is conceivable that the alignment slowly drifts during data collection.
The $176$ nm crystal (Fig.~\ref{ratevsfrequency}(c)) reveals a clear inhibition of the decay rate inside the band gap in good agreement with the calculated density of states.
Up to $12 \times$ inhibition is observed, which agrees with the results on the $170$ nm crystal.

\begin{figure}
\centering
\includegraphics[width=0.9\columnwidth]{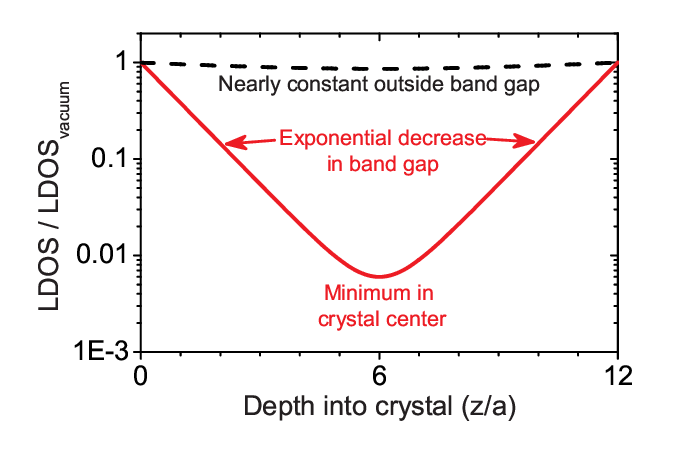}
\caption{(color online) Envelope of the LDOS normalized to vacuum versus depth into a finite band gap crystal with extent $L = 12 a$.
In the band gap the LDOS decreases exponentially away from the crystal surface to a minimum $< 10^{-2}$, as estimated from Ref.~\cite{Ishizaki09}.
Outside the gap the LDOS remains finite, as shown for a pseudogap as in Ref.~\cite{Lodahl04}.}
\label{DOScrystalsize}
\end{figure}

All theoretical studies predict strongly inhibited excited-state lifetimes in the range of the band gap~\cite{John90,Li01,Vats02,Wang03,Kristensen08}.
Indeed we observe drastically inhibited emission in our photonic band gap crystals.
Interestingly, the situation studied here leads to new physics since emission rates in real 3D photonic band gaps will reveal spatial inhomogeneity.
Based on theory for 2D crystals~\cite{Asatryan01} we expect the LDOS averaged over the unit cell to decrease exponentially with depth $z$ in the crystal at frequencies in the band gap, see Fig.~\ref{DOScrystalsize}.
In contrast, outside the gap the LDOS varies only slightly with $z$ from vacuum to the finite bulk value (\emph{cf.} Fig.~\ref{SEMbandstructureDOS}(c)).
In the gap, the LDOS is minimal at the crystal center with a minimum determined by the crystal size $L$~\cite{Ishizaki09}.
Hence we estimate the maximum inhibition at the center of our crystals to be more than $100 \times$, see Fig.~\ref{DOScrystalsize}.

Spatial inhomogeneity has intriguing consequences: experimentally we measure time-resolved emission that is proportional to the radiative rate $\gamma_{rad}$ times the LDOS-distribution~\cite{vanDriel07}.
Thus such an experiment is biased to quantum dots with a high emission rate and against dots with a strong inhibition.
This qualitatively explains why we observe a finite inhibition.
To take advantage of the strong inhibition deep inside 3D crystals calls for completely different experimental approaches, \emph{viz.}, the probing of excited-state populations to directly verify the theoretical predictions~\cite{John90,Li01,Vats02,Wang03,Kristensen08}.
We propose to use transient absorption as a probe of excited states that become more and more stable due to the shielding of the vacuum fluctuations in a 3D photonic band gap.

In addition to spatial inhomogeneity, non-zero non-radiative decay occurs for any real light source.
Hence, a time-resolved emission experiment yields a total rate that equals the sum of the radiative and non-radiative rates: $\gamma_{tot} = \gamma_{rad} + \gamma_{nrad}$.
When $\gamma_{rad}$ is strongly inhibited in a 3D photonic band gap, as is the case here, the rate $\gamma_{tot}$ is minimal with a lower bound given by $\gamma_{nrad}$.
While $\gamma_{nrad}$ is not well known for our quantum dots, we can estimate an upper bound from the maximum observed inhibition to be $1 / 12 \times 1.9 = 0.16 \mu s^{-1}$, which is reasonable in view of data on other colloidal dots~\cite{Leistikow09}.

Finally, it is well-known that the emission rate strongly depends on orientation~\cite{Vos09}.
Since our quantum dots are in suspension they sample all orientations of the transition dipole moment.
Thus the inhibition observed here is robust to orientation averaging.
This is in contrast to recently observed inhibited emission in 2D photonic crystals~\cite{Kress05}, and nanowires~\cite{Bleuse11} that concern dots with a particular orientation of the transition dipole moment.

We thank L\'{e}on Woldering, Hannie van den Broek, and Willem Tjerkstra for expert sample preparation.
This work is part of the research program of FOM that is financially supported by NWO.
WLV thanks NWO-Vici, and Smartmix Memphis.


\begin{thebibliography}{}
\bibitem{footnote:background}
We can conveniently distinguish the signal from background as signal counts are correlated with the laser pulses while background counts are uncorrelated and thus yield a constant level.
The background level was carefully determined by averaging data in the 6.4 ns time interval before the excitation pulse.
\end{thebibliography}
\end{document}